# High density array of epitaxial BiFeO$_3$ nanodots with robust and reversibly switchable topological domain states

Zhongwen Li,[1] Yujia Wang,[2] Guo Tian,[1] Peilian Li,[1] Lina Zhao,[1] Fengyuan Zhang,[1] Junxiang Yao,[1] Hua Fan,[1] Xiao Song,[1] Deyang Chen,[1] Zhen Fan,[1] Minghui Qin,[1] Min Zeng,[1] Zhang Zhang,[1] Xubing Lu,[1] Shejun Hu,[1] Chihou Lei,[3] Qingfeng Zhu,[4] Jiangyu Li,[4,5] Xingsen Gao,[1]* Jun-Ming Liu[6]*

**The exotic topological domains in ferroelectrics and multiferroics have attracted extensive interest in recent years due to their novel functionalities and potential applications in nanoelectronic devices. One of the key challenges for such applications is a realization of robust yet reversibly switchable nanoscale topological domain states with high density, wherein spontaneous topological structures can be individually addressed and controlled. This has been accomplished in our work using high density arrays of epitaxial BiFeO$_3$ (BFO) nanodots with lateral size as small as ~60 nm. We demonstrate various types of spontaneous topological domain structures, including center-convergent domains, center-divergent domains, and double-center domains, which are stable over sufficiently long time yet can be manipulated and reversibly switched by electric field. The formation mechanisms of these topological domain states, assisted by the accumulation of compensating charges on the surface, have also been revealed. These result demonstrated that these reversibly switchable**

[1]Institute for Advanced Materials and Guangdong Provincial Key Laboratory of Quantum Engineering and Quantum Materials, South China Normal University, Guangzhou 510006, China. [2]Shenyang National Laboratory for Materials Science, Institute of Metal Research, Chinese Academy of Sciences, 72 Wenhua Road, Shenyang 110016, China. [3]Department of Aerospace and Mechanical Engineering, Saint Louis University, Saint Louis, Missouri, 63103-1110, USA. [4]Shenzhen Key Laboratory of Nanobiomechanics, Shenzhen Institutes of Advanced Technology, Chinese Academy of Sciences, Shenzhen, Guangdong, 518055, China. [5]Department of Mechanical Engineering, University of Washington, Seattle, Washington 98195-2600, USA. [6]Laboratory of Solid State Microstructures and Innovation Center of Advanced Microstructures, Nanjing University, 21009, China.
*Corresponding authors. Email: xingsengao@scnu.edu.cn (X.S.G.) and liujm@nju.edu.cn (J.-M.L.).

topological domain arrays are promising for applications in high density nanoferroelectric devices such as nonvolatile memories

**INTRODUCTION**

Topological structures in ferroics have received substantial attention in recent years, and a number of exciting discoveries have been reported (*1-10*). Topological defects are usually considered as some singular regions of low dimensionalities, in which the order parameters cease to vary continuously (*2*). Two-dimensional topological defects include the well-known ferroic domain walls that have been extensively investigated for domain wall nanoelectronics (*1*). One-dimensional (1D) defects such as flux-closure vortex and skyrmion states, as schematically shown in fig. S1 in Supplementary Materials, have also been the focus of extensive researches (*3-7*), and specific center domain patterns were reported recently as well (*9*). These topological domain states, in combination with multiferroic functionalities, may lead to exciting new discoveries and device applications. For example, it was predicted that switchable polar vortex as small as 3.2 nm can remain stable, corresponding to an ultrahigh storage density of 60 Tbit/inch$^2$ (*3*).

While 1D topological defects in ferroelectrics and multiferroics have long been predicted, e.g. vortex domains by Naumov *et al* in 2004 (*3*), their experimental observations remain elusive in contrast to the well-studied analogs in ferromagnets (*1*). Recently, experimental evidences especially with flux-closure domain structures have gradually emerged, thanks to the powerful piezoresponse force microscopy (PFM) and advanced transmission electron microscopy (TEM) (*1, 4, 9-19*). For example, Nelson *et al* (*10*) and Jia *et al* (*11*) demonstrated the existence of half flux-closure quadrants in BiFeO$_3$ (BFO) and Pb(Zr,Ti)O$_3$ (PZT) thin films, respectively. Multi-state vortex-antivortex pairs were revealed in rare-earth manganites and other improper ferroelectrics by Cheong *et al* (*6, 14-16*). More recently, the periodic array of flux-closure vortices were found in PbTiO$_3$/SrTiO$_3$ multilayers by Tang *et al* (*12*) and in superlattices by Yadav *et al* (*13*). In BFO films, the formation of vortex and antivortex domain structures induced by electric field was also demonstrated by Balke *et al* (*17, 18*). In addition, specific center domain structures can be induced by radial electric field from charged scanning probe, as reported by Vasudevan *et al* (*9*) and Chen *et al* (*19*), although no spontaneously formed center domains have been observed. In spite of these tremendous progresses, robust yet switchable topological domain states critical for realizing future nanoelectronics remain elusive.

It is well known that polarization-strain coupling can be relieved in nanoscale ferroelectrics, resulting in more varieties of topological domains in confined systems (*20-23*). For example, unique domain quadrants and flux-closure quadrants were observed in micrometer sized single-crystal BaTiO$_3$ lamellae by Schilling *et al*, (*24*) McGilly *et al* (*25*), and McQuaid *et al* (*26*). Rodriguez *et al* reported evidence for vortex states in small PZT nanodot arrays (*27*). These studies suggested that dimension-reduction down to nanoscale is an effective strategy to manipulate topological defects, wherein delicate balances exist among various energetics such as exchanges, electro-elastic interaction, and electro-static interaction, which are closely related to dimension and surfaces. These findings could pave way towards nano-ferroelectronics based on local topological defects, if they are robust and switchable.

Inspired by this vision, we seek to explore possible topological defects (domain states) in high density array of epitaxial BFO nanodots. BFO is the most intensively studied multiferroic, promising for rich physics associated with various domain structures in addition to its superior electric properties (*28-30*). While vortex and center domain states generated by radial electric field were observed in BFO films, they cannot be spontaneously generated and their robustness and controllability remain to be seen (*9, 18, 19*). Here, we demonstrate individually controllable spontaneous center-type topological defects in high density BFO nanodot arrays, whose emergence does not require the assistance of any external electric field, a distinct advantage over earlier reports (*9, 19*). Using vector PFM and phase field simulation, we revealed a large percentage of center-type topological domains in individual nanodots, analogue to recently observed hedgehog spin meron states (*31*). These topological domain states are robust and individually controllable by electric field, a particularly promising characteristic, enabling on-demand manipulation for applications and offering opportunities for further exploring their novel properties in high density device applications.

## RESULTS

**Structural and PFM characterizations of BFO nanodot arrays**

The arrayed nanodots under investigation are roughly ~60 nm in lateral dimension and ~30 nm in height, corresponding to a pixel density of ~100 Gbit/inch$^2$. The fabrication procedure is schematically shown in fig. S2 in Supplementary Materials. To synthesize high quality nanodots, we employed a newly developed top-down ion-etching method using sacrificed nanoporous anodic alumina (AAO) template (*32*), which is different from conventional AAO template methods (*33-37*). The details of this synthesis can be found in the Materials and Method in Supplementary Materials. The SEM image (Fig. 1A) shows the well-ordered array of nanodots on SrTiO$_3$ substrate. The epitaxial structure of these nanodots was verified by XRD diffraction, featured by the (001) and (002) peaks in Fig. 1B. The out-of-plane lattice constant is ~4.03 Å, close to that of rhombohedral BFO films (*38, 39*).

The domain structures of these nanodots were characterized by PFM (*40*), noting that the vector PFM functionality allows the simultaneous mapping of vertical (out-of-plane) and lateral (in-plane) amplitudes (*V-amp.* & *L-amp.*) and phases (*V-pha.* & *L-pha.*) of piezo-response signals. The *V-amp.* and *V-pha.* images shown in Fig. 1 (C and D) for an as-prepared sample exhibit several unique features. The majority of nanodots have dark phase-contrast although some display bright-contrast, suggesting the favorable upward vertical polarization component. There are a small number of nanodots showing double-contrasts, implying the coexistence of upward and downward polarization components separated by a domain wall each case. More interesting is the lateral PFM image, and the typical examples are demonstrated in Fig. 1 (E and F), which cover the same area as Fig. 1 (C and D). Some characteristics can be highlighted here. First, the majority of nanodots have the half-dark and half-bright contrast in the *L-pha.* image (Fig. 1F). Besides, a small number of nanodots show even more complicated *L-pha.* contrast, e.g. dark/bright/dark or bright/dark/bright pattern from left to right. These features are verified by the presence of a coarse dark line in the *L-amp.* image (see Fig. 1D), exhibiting a domain wall like characteristic for later polarization component along the <010> -direction.

The PFM images for polarization switching induced by applying a scanning electric bias via the AFM probe, are also shown in fig. S3 in Supplementary Materials. Distinctly different contrasts in both the vertical and lateral PFM images can be found in the regions poled by the different DC bias voltages from +8 V and -8 V, respectively, indicating the concurrent switching of both the vertical and lateral polarization components (see fig. S3 (A-D) in Supplementary Materials). Moreover, the point-wise polarization reversal was checked on a number of nanodots one by one and the typical amplitude and phase loops are plotted in fig. S3 (E and F) in Supplementary Materials. The square-like phase loop evidences a 180° phase difference, indicating a complete switching of the point-wise vertical polarization with a coercive voltage of ~ 4 V.

From these PFM images, we can identify different domains, as shown in the enlarged images in the gap between two columns of the PFM images in Fig. 1. The first type (Type-I) shows the dark/bright contrast in the *L-pha.* image but the uniform dark contrast in the *V-pha.* image. The second (Type-II) exhibits the bright/dark contrast in the *L-pha.* image and uniform bright contrast in the *V-pha.* image, a reversed contrast with respect to Type-I. Type-III shows the dark/bright/dark contrast in the *L-pha.* image together with the dark/bright double-contrasts in the *V-pha.* image. Type-IV shows the reverse contrast to Type-III, with the bright/dark/bright contrast in the *L-pha.* image and bright/dark double-contrasts in the *V-pha.* image. The observations thus identify an emergent phenomenon and also raise an important question: what is the polarization distribution in these domains?

**Domain structure reconstruction**

The above analysis on the lateral PFM images can capture the polarization components along the *x*-axis and *z*-axis. Given the fact that the polarization may point to any directions in three-dimensional (3D) space, a full determination of polarization distribution in an individual nanodot must consult to additional data regarding the *y*-axis component. This can be performed by combining the PFM data for different in-plane rotational angles of the sample (*40-46*). Taking the data in Fig. 1 (C to F) as reference where the sample alignment angle is set as 0°, we conducted additional PFM imaging by rotating the sample clockwise for a set of given angles. The complete PFM data for an array of nanodots and the selected individual nanodots at different rotation angles can be found in Fig. S4 & S5 in Supplementary Materials. Clearly, both the topographic and vector PFM images for the same region before and after the rotations can match well without apparent image distortion. To resolve the 3D polarization distribution, we conducted a vector PFM mapping following the method proposed by Rodriguez *et al* (*44*), Gruverman and Kalinin (*45*).

We first chose a nanodot with a typical Type-I PFM contrast as a paradigm, and the construction procedure is demonstrated in Fig. 2. The PFM images (amplitude and phase) for the *x*-component (lateral PFM at angle of 0°, with cantilever parallel to the <010> crystallographic direction), *y*-component (lateral-PFM at angle of 90°, for sample clockwise rotation for 90° with cantilever parallel to the <100> direction), and *z*-component (vertical PFM), are presented respectively in Fig. 2 (A to C). Then the amplitude and phase images can be converted to the PFM piezoresponse signal contours of *x*-component (*PFMx*), *y*-component (*PFMy*), and *z*-component (*PFMz*), according to function $R\cos(\theta)$ (*R* is the amplitude, and $\theta$ is the phase angle), as shown in Fig. 2 (D to G). Finally, the as-evaluated *PFMx* and *PFMy* images are converted into the 2D vector contours using the MATLAB programing, as presented in Fig. 2H for the amplitude map and Fig. 2I for the vector angle map, reflecting the local polarization distributions for in-plane components. The 3D vector contours can also be constructed from the *PFMx, PFMy*, and *PFMz* data, as demonstrated in Fig. 2, J and K. From the resulting contours, we are able to identify that this nanodot possesses a unique domain structure with all the lateral polarization vectors pointing radially to the center, as schematically illustrated in Fig. 2L, where the vertical components over the whole nanodot are pointing downwards. That is the so-called center type topological domain, as defined by Mermin (*2*). Such a domain state also matches well with the PFM contrasts obtained after arbitrary angle rotations, for example by 45°. This domain structure is somewhat similar to the quadrant center domain in previously reported by Vasudevan *et al*, created by a large radial electric field in the BFO film via AFM tip (*9*). Notably, our domain structure has its polarizations rotating more or less continuously, indicating a roughly isotropic center domain analogue to hedgehog spin topological Meron states (*31*). More importantly, this specific domain pattern was generated spontaneously in the as-fabricated nanodots without any assistance of external field. This is a major step towards practical applications, drastically different from the quadrant ones in previous reports (*9*, *19*). The spontaneous occurrence of the center topological domains in the virgin state also implies that they are probably most stable states, essential for further manipulation and application as a functional unit in devices.

Similar analysis has been conducted on other types of nanodots, and the results are summarized in Fig. 3, with more detailed PFM data presented in fig. S5 in Supplementary Materials. Now it is understood that the Type-I (Fig. 3A) is the radially center-convergent domain with polarizations pointing inward the center. The Type-II (Fig. 3B) shows the uniform dark contrast in vertical PFM image and dark/bright contrast in the lateral PFM image, which turns out to be a radial center-divergent domain with radial outward polarizations. The Type-III (Fig. 3C) shows a double-contrast variation in the vertical PFM image, as well a triple-contrast variation (bright/dark/bright) in the lateral image for angle 0° and a quadrat-domain contrast variation for angle 90°, corresponding to a double-center domain structure consisting of both a center-convergent domain and a divergent one. The Type-IV (Fig. 3D) is the reverse case of the Type-III and it is also a double-center domain structure consisting of a center-divergent domain and a convergent one, which can be considered as an equivalent state to Type-III. As a result, these domain structures can be re-classified into three types: center-convergent domain, center-divergent (convergent-reverse) domain, and double-center domain.

To complement the above analysis, we counted the percentages of different types of domain structures in an array of totally 238 nanodots at virgin state without any electric poling. It was found that 61% of them are the radial center-convergent domains, ~14% are the radial center-divergent domains, ~24% are the double-center domains, and the rest (~1%) are other types of complex domains.

## DISCUSSION

The observations of these center domains instead of earlier reported quadrant center domains is rather surprising, since the polarizations deviate from any of the predetermined eight <111> equivalent orientations for BFO. The formation of these domains is probably driven by the competitions among depolarization energy, polarization-strain coupling, and surface strain, all of which can greatly change the local anisotropy and thus the polarization distribution. Besides that, the possible non-uniform strain in the nanodots can generate the flexo-electric rotations that can drive the polarization away from their original directions (*47, 48*). In addition, the three types of center domain structures all possess the head-to-head or tail-to-tail charge cores. This generates potentially high electrostatic energy, and one possible source favoring these structures is the surface and edge effects in low dimensional systems, as predicted by Hong *et al* using the first principles simulation (*49*). However, our system is much bigger (~60 nm) than the predicted one (a few unit cells in diameter), and the surface or edge effects may not provide sufficient formation energy for center domains. Another possible reason is associated with the charge accumulation on the top surface, and this mechanism is supported by our phase field simulation and additional experiments.

To mimic the experimental conditions, the nanodot was modeled as a nanoscale cylinder with a diameter of 64 nm and height of 30 nm (Fig. 3E) and simulated by phase-field simulation, and the simulation details can be found in Supplementary Materials. Fig. 3 (F to H) show the simulated polarization vector contours of three BFO nanodots imposed with three different charge distribution states at the top surface respectively: uniform positive charge (Fig. 3F), uniform negative charge (Fig. 3G), and combination of both positive and negative charges (Fig. 3H), resulting in different center domain states. It was found that the negative charges favor upward vertical polarizations with all in-plane polarizations inwards, showing a center-convergent domain structures. For the positive charge state, the polarization distribution shows a center-divergent domain structure. In the case of half-positive and half-negative charge state, the nanodot shows two center domains, i.e. a double-center domain structure. All the three domain structures match well with topological domain states observed in the experiment.

The observed head-to-head or tail-to-tail charge domain cores are somewhat analogous to charge domain walls, which can be stabilized by the charge accumulation from interior electrons, holes, ionic defects, or exterior absorbed charges adjacent to the walls, as summarized in a previous book chapter by Seidel (*7*). In our case, these nanodots were fabricated from the BFO films deposited at a relatively low oxygen pressure (~3.0 Pa), and thus contained rather high-density oxygen vacancies and other charged carriers. As a result, the charge carriers can emigrate to the charge cores or domain heads/tails to stabilize the charge domain walls/cores. The free electrons from the ambient can also be injected onto the surface for the charge domain walls/cores, reducing the formation energy. To visualize the possible charges on the nanodots, we carried out the scanning thermo-ionic microscopy (STIM) studies recently developed to map the distribution of ionic and other species as charge carriers (*50*), as detailed in the Supplementary Materials. One example is the STIM amplitude mapping shown in Fig. 3I, superimposed on the 3D topology. It was observed that the STIM contrast is not uniform, with some dots showing bright contrast at the centers, whereas others exhibiting dark contrast at the center together with bright contrast at the outer edges forming ring-like patterns. The former reflects the accumulation of mobile ions (likely oxygen vacancies in our cases) at the top center surface carrying positive charges, and the latter may be due to the accumulation of electron charges (dark contrast) at the top center and oxygen vacancies (bright contrast) at the outer edge or bottom of the dots. Such non-uniform accumulation of charges may act as the major driving force for the center domains. This argument is also supported by the occurrence of plenty of spontaneous charged domain walls in the oxygen deficient BFO films prior to the ion beam etching (see Supplementary Materials fig. S6 (A to C). In contrast, typical stripe-like domain patterns free of charged domain walls (e.g. 71° domain walls) are dominant in the BFO films deposited at much higher oxygen pressure of ~15 Pa (see Supplementary Materials fig. S6 (D to F). Once the low oxygen-deficient film with stripe domains was etched into similar nanodots, one can see from the PFM data (see Supplementary Materials fig. S7) that most of the nanodots still exhibit the stripe-like domain patterns without any charge domain walls. These observations indicated that the mobile charge carriers play critical role in the non-uniform charge accumulations and provide a driving force for the formation of charge domain walls or the center topological domains. This does create an opportunity for tailoring the different topological defects (e.g. center domain, vortex states) by properly tuning the charge accumulation level and geometric parameters of nanoferroelectrics.

Equally exciting is that these center domains can be reversibly switched and verse vice. Fig. 4 (A to D) shows the *V-pha.* and *L-pha.* images for a nanodots array, in which three of the individual nanodots were subjected to electric pulses by fixing the AFM probe on the selected individual nanodots. Upon applying an electric pulse of +8 V, one can clearly see that the domain can be switched into center-divergent domain, while applying a pulse of -8 V converts the center-divergent domain to the center-convergent domain, as indicated by the complete reversal of contrast patterns in both the *V-pha.* and *L-pha.* images for the selected nanodots (see Fig. 4 (B and C)). These reversed center domains can also be switched back to their initial states by applying another set of electric pulses with opposite polar (see Fig. 4 (C and D)), indicating that the switching is reversible and individually addressable. The domain switching behavior for an array of nanodots triggered by scanning electric bias, and their retention properties are shown in Fig. 4 (E to H). It was found that applying two different sets of scanning bias voltages (±8 V) via the AFM probe on an array of nanodots produces two distinct regions with rather uniform center-convergent domains (with upward polarization) for +8 V and center-divergent (with upward polarization) for -8 V, respectively. After retention duration of 6000 min, the center domains remained nearly unchanged, except one at the border which was switched from center-convergent back to divergent domain. After even longer duration (24000 min), most of the topologic center domains remain unchanged. This suggests that the center domains are rather robust in ambient yet reversibly switchable under electric field. This also enables further manipulation of such topological domain states by electric field, promising for applications in high density devices, e.g. nonvolatile memories.

In summary, we have observed different types of spontaneous ferroelectric topological domain states in multiferroic BFO nanodots array. The domain configurations in the nanodots array were examined by vector PFM analysis, which reveals the existence of center-convergent domain, reverse center (center-divergent) domain, as well as double-center domains. These domain structures exist spontaneously in the as-deposited virgin states, stabilized by the accumulation of charge on the top surface, as supported by our phase field simulations and STIM characterization. Furthermore, these topologic domains are rather stable and can be effectively and reversibly switched by electric fields, which is promising for potential applications in high density memory devices.

## MATERIALS AND METHODS

### Fabrication of BFO nanodot array

The nanodot arrays were fabricated by a sacrificed ultrathin (~300 nm) AAO template etching technique, and the details can be found in our previous work (37). The simplified procedure is shown Fig. S2. In brief, an epitaxial BFO film (~100 nm thick) with a ~50 nm $SrRuO_3$ bottom electrode layer was first grown on a (001)-oriented $SrTiO_3$ (STO) substrate by pulsed laser deposition (PLD). Then an ultrathin (~300 nm) AAO membrane mask (refer to the Supplementary Materials) was transferred onto the BFO film surface, followed by the $Ar^+$ ion beam etching. The ion beam gradually changed the shapes of both mask and film, and developed an epitaxial nanodots array. Finally the mask was removed by mechanical method and the attractively nanostructures were obtained. The as-prepared nanodots maintain both epitaxial structures and good electrical properties close to their parent films. This procedure is reproducible and the as-prepared samples for subsequent characterizations show nice consistency with each other.

### Structure, PFM, and STIM characterizations

The topography images of the as-prepared nanodots arrays were measured by AFM at contact mode (AFM, Cypher Asylum Research). The crystallinity of nanodots was characterized by X-ray diffraction (PANalytical X′ Pert PRO). The ferroelectric domain structures of these nanodots were characterized by piezoresponse force microscopy (PFM, Asylum Cypher) using conductive probes (EFM arrow, Nanoworld). The local piezoresponse loop measurements were carried out by fixing the PFM probe on a selected nanodot, and then applying a triangle-square waveform accompanying with ac driven voltage, via the conductive PFM probe. To improve the PFM sensitivity, we adopted a dual frequency resonant-tracking technique (DART) also provided by Asylum Research (*36*). The vector PFM function of our AFM unit allows simultaneously mapping of the vertical (out-of-plane) and lateral (in-plane) amplitude (*V-amp* & *L-amp*) and phase (*V-pha* & *L-pha*) signals from the nanodots one by one. To determine 3D domain structures, both the vertical and lateral PFM images were conducted for different in-plane sample rotation angles. In proceeding, we marked the sample before the rotations so that the same scanned area can be tracked. The implementation of STIM (*50*) here utilizes an Asylum Research MFP-3D AFM equipped with Anasys ThermaLever AN2-300 thermal probe, and the technical details can also be found in Fig. S8 in Supplementary Materials.

## Phase field simulation details

We simulated the nanodot as a cylinder surrounded by vacuum, as shown in Fig. 5E. We chose a mesh size of $x_0 = y_0 = 2$ nm and $z_0 = 0.6$ nm so that a $64 \times 64 \times 50$ mesh corresponds to a $128 \times 128 \times 30$ nm$^3$ model. The diameter of the cylinder for the nanodot is 64 nm and the height of 30 nm inside the simulation mesh. At the bottom and top of the nanodot, the short and open circuit boundary conditions are applied, respectively. This type of electric boundary condition is consistent with the experimental setup that the nanodot is grown on a bottom electrode. To simulate the different charge distribution states, the charge is applied on the top surface of the nanodots for all the three different states: uniform positive charge, uniform negative charge, and combination of positive and negative charges. The evolution of polarization in BFO nanodots is studied by phase field simulations, which are based on the time-dependent Ginzburg-Landau equation. The Landau-Devonshire energy of BFO is expanded as a fourth order polynomial using the parameters adopted from previous literatures (*51*, *52*). More simulation details can also be found in the Supplementary Materials.

## SUPPLEMENTARY MATERIALS

Supplementary materials for this article is available at xxx

Nanoporous anodic alumina (AAO) membrane masks fabrication

Scanning thermo-ionic microscopy (STIM)

Phase field simulation details and parameters fig. S1. Schematic diagrams for different 1D topological defects.

fig. S2. Schematic procedures for the fabrication and PFM characterization of BFO nanodots array sample.

fig. S3. PFM images for the nanodot array after poling by scanning bias voltages of ±8 V.

fig. S4. 3D PFM images for a nanodot array.

fig. S5. Single dot PFM images for four typical topologic domains.

fig. S6. A comparison of domain structures between an oxygen deficient BFO film and a less oxygen deficient film.

fig. S7. Stripe domain structures in nanodots derived from a less oxygen deficient BFO film.

fig. S8. Scanning thermal-ionic microscopy (STIM) images of a nanodot array in this work.

**Acknowledgments:** The authors would like to thank the National Key Research and Development Program of China (Nos. 2016YFA0201002 & 2016YFA0300101), the State Key Program for Basic Researches of China (No. 2015CB921202), the Natural Science Foundation of China (Nos. 11674108, 51272078, 51431006, 51401212,), the Project for Guangdong Province Universities and Colleges Pearl River Scholar Funded Scheme (2014), Science and Technology Planning Project of Guangdong Province (No. 2015B090927006), the Natural Science Foundation of Guangdong Province (No. 2016A030308019).

**Author Contributions:** X.S.G. conceived and designed the experiments. Z.W.L. conducted the experiments. Z.F, G.T., F.H., D.Y.C. and L.N.Z. contributed to the sample fabrications and XRD measurements. Z.W.L., F.Y.Z., X.S., P.L.L., Z.F. Q.F.Z, and J.X.Y. carried out the PFM measurements. J.X.Y. and Z.Z. contributed to the AAO preparation. S.J.W., X.B.L., M.H.Q., M.Z. and S.J.H. contributed to the data interpretation. Y.J.W., C.H.L. and J.Y.L conducted to the phase field simulation. X.S.G. & J.-M.L. conducted the data interpretation and co-wrote the article. All authors discussed the results and commented on the manuscript.

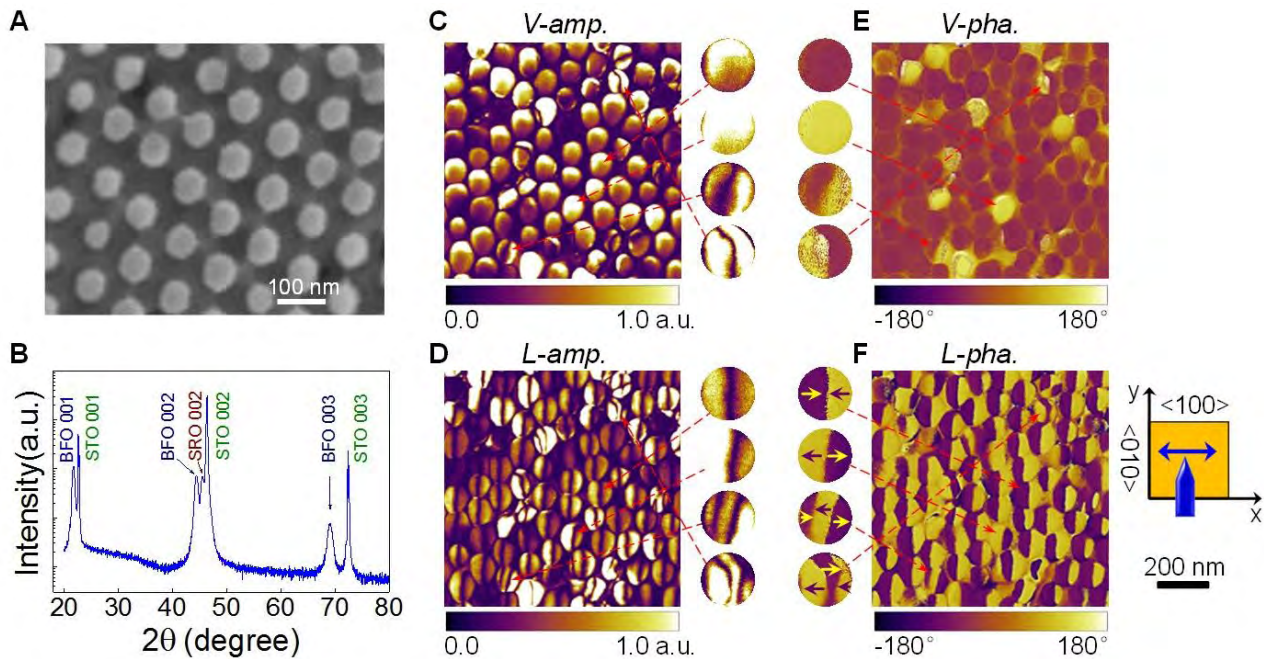

**Fig. 1. Structure and PFM images for a BFO nanodot array.** (**A**) Scanning electron microscopy (SEM) image. (**B**) XRD diffraction pattern. (**C** and **D**) Vertical PFM amplitude (**C**) and phase (**D**) images for the as prepared nanodot sample (The labels *V-amp.* and *V-pha.* present the vertical PFM amplitude and phase, respectively). (**E** and **F**) Lateral PFM amplitude (**E**) and phase (**F**) images (*L-amp.* refers lateral PFM amplitude and *L-pha.* refers lateral PFM phase). Some single nanodot PFM images are zoomed-in and shown in the insets in the gap between **C**, **D** and **E**, **F** illustrating some typical PFM contrast variants frequently observed in the nanodots. The inset schematic diagram of cantilever indicates that the cantilever is parallel to the y-axle (<010> direction), thus the contrasts in the lateral PFM image reflect the x-components of polarization vectors (along <100> direction).

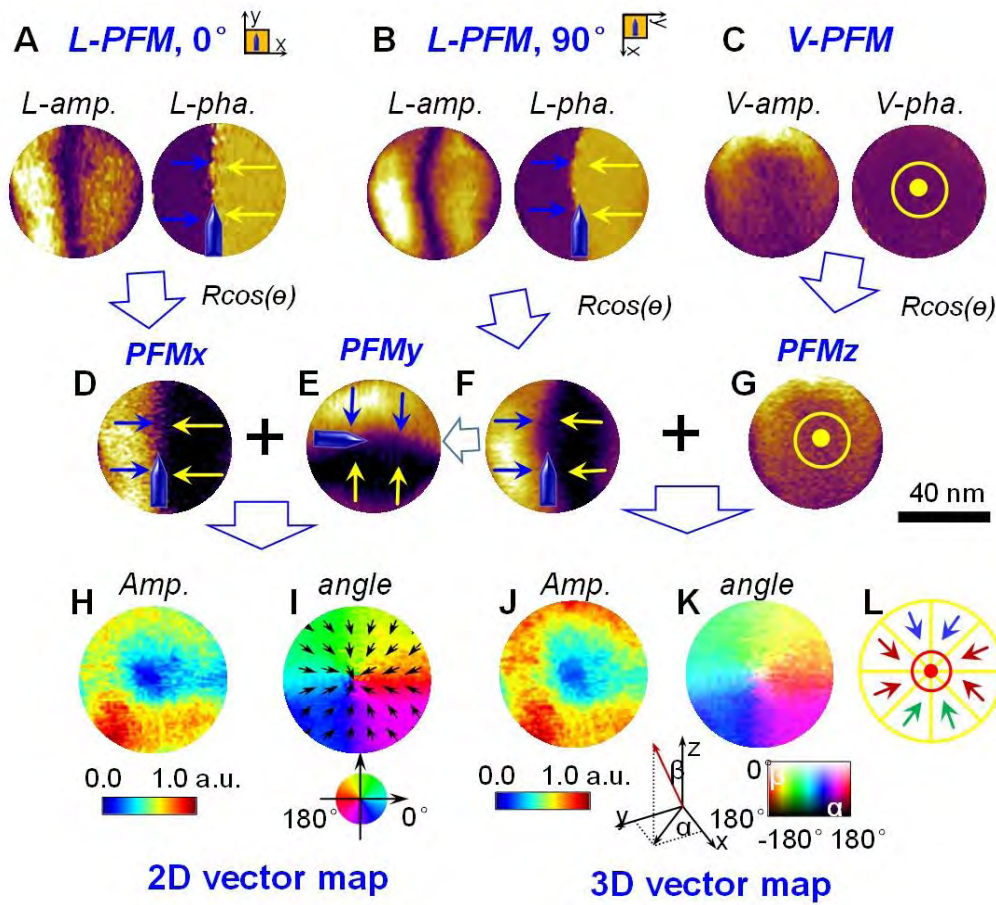

**Fig. 2. 3D domain reconstruction using vector PFM analysis for a typical nano-domain inside a single nanodot.** (**A** to **C**) The lateral PFM amplitude and phase images with sample rotation for 0º (**A**) and 90º (**C**), along with the vertical PFM images (**C**). (**D** to **G**) The piezoreponse images (PFM) for both *x*-, *y*- and *z*- components derived from the combination of amplitude and phase images according to *Rcos(θ)* (*R*: amplitude, *θ:* phase angle), in which the *x*-components of PFM (**D**, *PFMx*) is converted from **A**, **F** from **B**, and *z*-component of PFM (**G**, *PFMz*) from **C**. The *y*-component of PFM (E, *PFMy)* is obtained by anticlockwise rotating the image F for 90º, so that the local PFM signals in **D**, **E**, and **G** are from the same locations. (**H** and **I**) The 2D vector contour maps including amplitude map (**H**) and phase angle map (**I**) converted from the combination of *PFMx* (**D**) *PFMy* (**E**), presenting the lateral polarization distributions. (**J** and **K**) The 3D vector map including amplitude map (**J**) and angle map (**K**), derived from the combination of *PFMx* (**D**), *PFMy* (**E**), and *PFMz* (**G**). (**L**) The schematic diagram for the 3D domain structures.

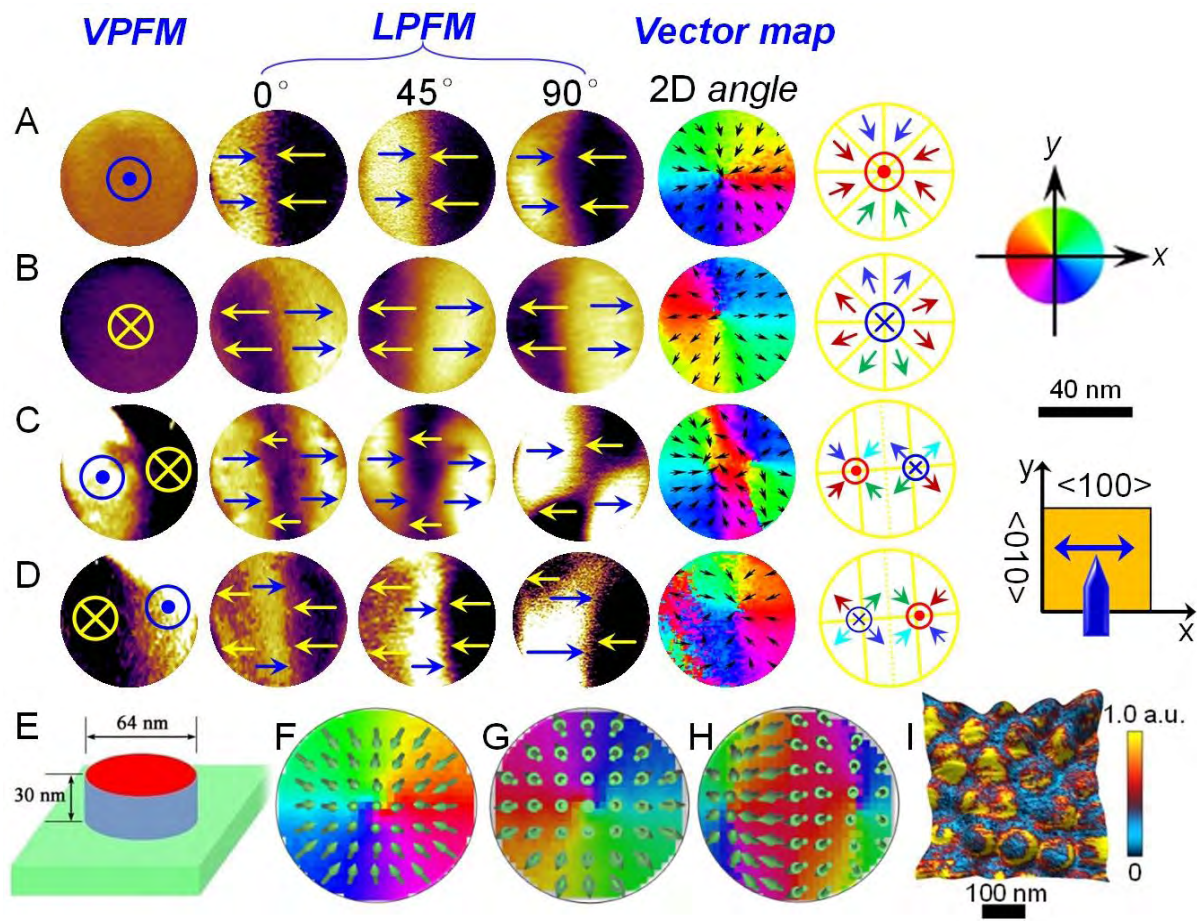

**Fig. 3. Vector PFM images and vector maps, along with the simulated contours for some typical topological domains in the nanodots.** (**A** to **D**) PFM images for the four typical domain structures frequently observed in the nanodots, which are center-convergent domain (**A**), center-divergent domain (**B**), double-center domain (**C**), and reverse double-center domain (**D**) that is equivalent to the double-center domain in **C**. The micrographs from left to right are, piezoresponse images derived from $Rcos(\theta)$ ($R$: the amplitude, $\theta$: the phase angle) for vertical PFM ($PFMz$), and lateral PFM for three sample ration angle of 0° ($PFMx$), 45° ($LPFM$ along <110> direction), and 90° ($PFMy$), respectively, as well as their corresponding 2D vector angle maps and schematic domain configurations. (**E**) The cylinder model for phase-field simulation of the center domains in nandots. (**F** to **H**) the three different polar vector contour maps derived from the simulation of different charge distribution states: positive charge (**F**), negative charge (**G**), and half-positive and half-negative charge (**H**) states. The arrows in the simulated vector contours present the microscopic polarization directions, and the color scales showing the angular distribution of lateral polarization. (**I**) The scanning thermal ionic microscopy (STIM) amplitude contrast map superimposed in 3D surface topology, indicating the non-uniform accumulation of mobile ionic charges.

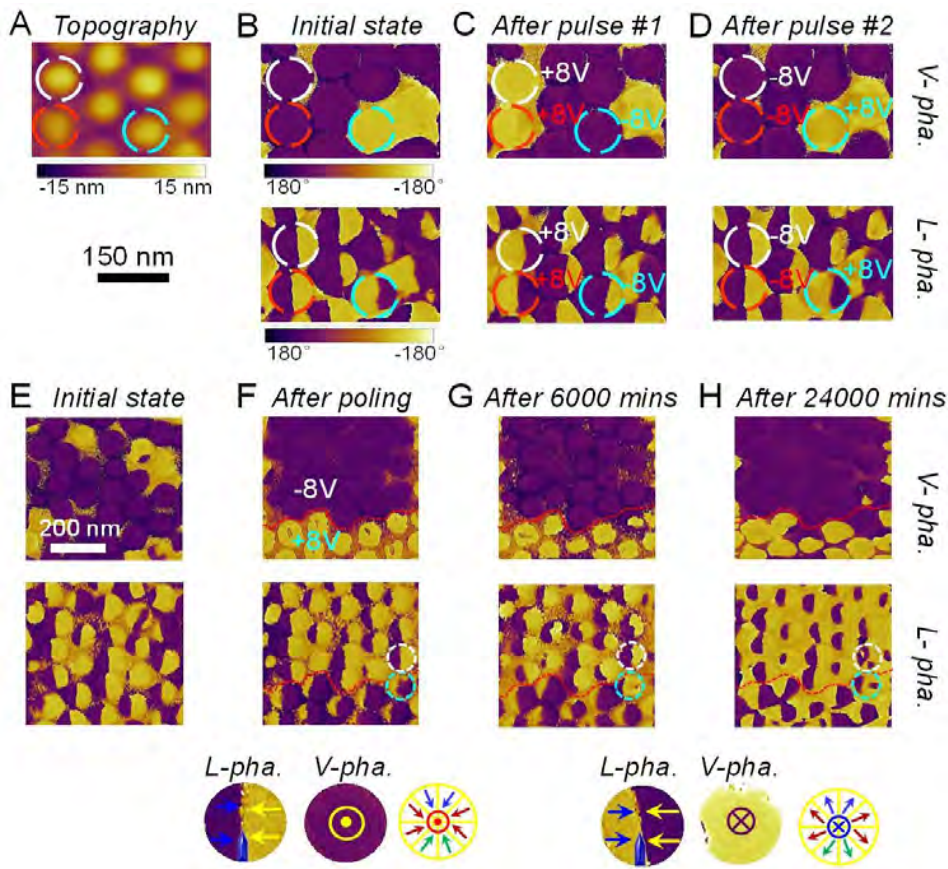

**Fig. 4. Electric switching behaviors and retention properties for an array of center domains.** (**A** to **D**) The topology and PFM images for a nanodot array illustrating the reversible switching of center domains triggered by applying pulsed electric fields on individual dots, including the topological image (**A**), and the PFM vertical (upper panel) and lateral (lower panel) phase images at initial state (**B**), after the first set of electric pulses (**C**), and after second set of electric pulses (**D**). (**E** to **H**) The PFM images showing the electric switching of center domains by scanning electric bias and the retention properties for the switched center domains, for the initial state (**E**), the just switching state (**F**), after a retention duration for 6000 minutes (**G**), and after an elongated retention duration for 24000 minutes (**H**). The circles in **G** and **H** indicate the contrast changes in certain nanodots during retention test. The insets below the panels illustrate the characteristic contrasts in *L-pha.* and *V-pha.* images for center-convergent (left) and center-divergent (right) domains, respectively.

Supplementary Materials for

# High density array of epitaxial BiFeO$_3$ nanodots with robust and reversibly switchable topological domain states

Zhongwen Li, Yujia Wang, Guo Tian, Peilian Li, Lina Zhao, Fengyuan Zhang, Junxiang Yao, Hua Fan, Xiao Song, Deyang Chen, Zhen Fan, Minghui Qin, Min Zeng, Zhang Zhang, Xubing Lu, Shejun Hu, Chihou Lei, Qingfeng Zhu, Jiangyu Li, Xingsen Gao*, Jun-Ming Liu*

**This PDF file includes:**

- Nanoporous anodic alumina (AAO) membrane masks fabrication
- Scanning thermo-ionic microscopy (STIM)
- Phase field simulation details and parameters
- fig. S1. Schematic diagrams for different 1D topological defects.
- fig. S2. Schematic procedures for the fabrication and PFM characterization of BFO nanodots array sample.
- fig. S3. PFM images for the nanodot array after poling by scanning bias voltages of ±8 V.
- fig. S4. 3D PFM images for a nanodot array.
- fig. S5. Single dot PFM images for four typical topologic domains.
- fig. S6. A comparison of domain structures between an oxygen deficient BFO film and a less oxygen deficient film.
- fig. S7. Stripe domain structures in nanodots derived from a less oxygen deficient BFO film.
- fig. S8. Scanning thermal-ionic microscopy (STIM) images of a nanodot array in this work.

**Nanoporous anodic alumina (AAO) membrane masks fabrication.** The ultrathin AAO masks were fabricated by a two-step anodization process. The first anodization of Al sheet was carried out in 0.3 M of H$_2$C$_2$O$_4$ solution for 24 hours at 5 $^o$C, then the anodized Al sheet was completely removed in mixture of H$_3$PO$_4$ and CrO$_3$ (6.0 wt% and 1.8 wt%) at 45$^o$C for 12 hours. The second anodization was carried out for 5 mins at 5 $^o$C to get well ordered pores, which was followed by an etching in CuCl$_2$ at 10 $^o$C to detach the alumina layer from the Al sheet. In sequence, the alumina barrier layer was removed during the pore widening process in 5 wt% $H_3PO_4$ at 35 $^oC$ for 35 min. Finally, we obtained ~300 nm-thick AAO ultrathin membranes.

**Scanning thermo-ionic microscopy (STIM).** In this work, a newly developed scanning thermo-ionic microscopy (STIM) technique was employed to probe the local thermal ionic activity at nanoscale, based on imaging of Vegard strain induced by ionic species. The details of STIM can be found in previous report. The basic principle is based on Vegard strain induced by concentration changes, as well its converse effect that diffusion of ionic and electronic species can be driven by the gradients in hydrostatic stress. Once the sample is imposed with oscillating heating by passing an AC current through the micro-fabricated resistor localized on top of the tip, it will generate an oscillation in local temperature. Such local temperature oscillation in turn produces a concentrated thermal expansion strain and thus the thermal stress at the second harmonic. This resulted in the second harmonic response of the thermally induced cantilever vibration associated with thermal expansion present in all solids, as well as the fourth harmonic response caused by local transport of mobile species, which is only sensitive to ionic activities.

The implementation of STIM here utilizes an Asylum Research MFP-3D AFM equipped with Anasys ThermaLever AN2-300 thermal probe. The driven force is inputted from the ac current at the cantilever sample contact resonance frequency of $f_0$ first, and then the output thermal probe vibration is measured at $f_0/2$ and $f_0/4$, respectively, using lock-in to enhance the sensitivity. These give us the second and fourth harmonic responses that correspond to local thermomechanical and thermosionic activities, respectively (see fig. S8).

**Phase field simulation details and parameters.** The evolution of polarization in BFO nanodots was studied by phase field simulations, which are based on the time-dependent Ginzburg-Landau equation. The order parameters are chosen as the three components of spontaneous polarizations. We simulated the nanodot as a cylinder on conductive bottom electrode which are surrounded by vacuum. We chose mesh sized of $x_0 = y_0 = 2$ nm and $z_0 = 0.6$ nm so that a 64 × 64 × 50 mesh corresponds to a 128 × 128 × 30 nm$^3$ model. The diameter of the nanodot is 64 nm and the height 30 nm inside the square mesh. At the cylindrical surface, we ignore the surface effect by setting the polarization components outside the cylinder to be zero. At the bottom and top of the nanodot, the short and open circuit boundary conditions are applied, respectively. This type of electric boundary condition is consistent with the experimental setup that the nanodot is grown on an electrode and the top surface is contacted with vacuum.

The charges are incorporated into this system as a sheet of charge density σ on the top surface of the nanodot. The electrostatic driving force is obtained by solving the Maxwell equation, $\nabla \cdot D = \rho$, where, ρ is calculated as σ/$z_0$, assuming the surface charge occupy one grid point thickness. For the case of half-positive-half-negative, we applied two semicircular charge sheets of different sign on the top surface, in which the separation between the two sheets is 2 nm. In this work, the sheet charge density is set as of 2C/m$^2$ during the simulation.

3D phase field models for BiFeO$_3$ nanodots are constructed using the three components of the polarization vectors as order parameters. The Helmholtz free energy was considered as the system energy function:

$$f = f_{bulk}(P_i) + f_{grad}(P_{i,j}) + f_{elas}(P_i, \varepsilon_{ij}) + f_{elec}(P_i, E_i) \tag{1}$$

The first term is the bulk energy or Landau-Devonshire energy:

$$f_{bulk} = \alpha_1(P_1^2 + P_2^2 + P_3^2) + \alpha_{11}(P_1^4 + P_2^4 + P_3^4) + \alpha_{12}(P_1^2 P_2^2 + P_2^2 P_3^2 + P_3^2 P_1^2) \tag{2}$$

The second term is the gradient energy or Ginzburg energy:

$$F_{grad} = \frac{1}{2} G_{11}(P_{1,1}^2 + P_{2,2}^2 + P_{3,3}^2) + G_{12}(P_{1,1} P_{2,2} + P_{2,2} P_{3,3} + P_{3,3} P_{1,1})$$
$$+ \frac{1}{2} G_{44}\left[(P_{1,2} + P_{2,1})^2 + (P_{2,3} + P_{3,2})^2 + (P_{3,1} + P_{1,3})^2\right] \tag{3}$$
$$+ \frac{1}{2} G'_{44}\left[(P_{1,2} - P_{2,1})^2 + (P_{2,3} - P_{3,2})^2 + (P_{3,1} - P_{1,3})^2\right]$$

The third term is the elastic energy:

$$f_{elas} = \frac{1}{2} C_{ijkl}(\varepsilon_{ij} - \varepsilon_{ij}^o)(\varepsilon_{kl} - \varepsilon_{kl}^o) \tag{4}$$

where, $\varepsilon_{ij}$ is the total strain and $\varepsilon_{ij}^o$ is the spontaneous strain due to the paraelectric/ ferroelectric phase transition. Their difference is the elastic strain. The spontaneous strain is related to the polarization by the electrostrictive coefficients:

$$\varepsilon_{11}^o = Q_{11}P_1^2 + Q_{12}(P_2^2 + P_3^2)$$
$$\varepsilon_{22}^o = Q_{11}P_2^2 + Q_{12}(P_3^2 + P_1^2)$$
$$\varepsilon_{33}^o = Q_{11}P_3^2 + Q_{12}(P_1^2 + P_2^2) \quad (5)$$
$$\varepsilon_{23}^o = Q_{44}P_2P_3$$
$$\varepsilon_{13}^o = Q_{44}P_1P_3$$
$$\varepsilon_{12}^o = Q_{44}P_1P_2$$

The last term is the electrostatic energy:

$$f_{elec} = -\frac{1}{2}E_i(\varepsilon_0\varepsilon_r E_i + P_i) \quad (6)$$

where, $\varepsilon_0$ is the permittivity of vacuum and $\varepsilon_r$ is the background relative dielectric constant.

Following previous work, we assume that the equilibrium of mechanical stress and electrical field is much faster than the evolution of domain structures. Thus, for each polarization configuration, the mechanical and electrical equilibrium equations

$$\sigma_{ij,j} = 0 \quad (7)$$
$$D_{i,i} = 0 \quad (8)$$

are solved to obtain the corresponding driving forces.

The evolution of polarizations is simulated by the time-dependent Ginzburg-Landau equation:

$$\frac{dP_i}{dt} = -L\frac{\delta f}{\delta P_i} \quad (9)$$

We use the backward Euler methods to trace the evolution of polarizations.

All coefficients of BiFeO$_3$ are adopted from previous literatures and are listed in Table S1. These coefficients are normalized by the following formulae:

$$\alpha_1^* = \alpha_1/\alpha_0 \quad \alpha_{11}^* = \alpha_{11}P_0^2/\alpha_0 \quad \alpha_{12}^* = \alpha_{12}P_0^2/\alpha_0$$
$$g_{11}^* = G_{11}/G_{110} \quad g_{12}^* = G_{12}/G_{110} \quad g_{44}^* = G_{44}/G_{110} \quad g_{44}'^* = G_{44}'/G_{110}$$
$$C_{11}^* = C_{11}/\alpha_0 P_0^2 \quad C_{12}^* = C_{12}/\alpha_0 P_0^2 \quad C_{44}^* = C_{44}/\alpha_0 P_0^2$$
$$q_{11}^* = Q_{11}P_0^2 \quad q_{12}^* = Q_{12}P_0^2 \quad q_{44}^* = Q_{44}P_0^2$$

Where, $\alpha_0 = 3.945 \times 10^8 \ C^{-2} \cdot m^2 \cdot N$, $P_0 = 0.63 \ C \cdot m^{-2}$, $G_{110} = 1.578 \times 10^{-9} C^{-2} \cdot m^4 \cdot N$

**Table S1. The coefficients used in the phase field simulations.**

| Coefficient | Value | Unit | Normalized coefficient | Value |
|---|---|---|---|---|
| $\alpha_1$ | $-3.945 \times 10^8$ | $C^{-2}m^2N$ | $\alpha_1^*$ | -1 |
| $\alpha_{11}$ | $5.42 \times 10^8$ | $C^{-4}m^6N$ | $\alpha_{11}^*$ | 0.545 |
| $\alpha_{12}$ | $1.54 \times 10^8$ | $C^{-4}m^6N$ | $\alpha_{12}^*$ | 0.155 |
| $G_{11}$ | $9.468 \times 10^{-10}$ | $C^{-2}m^4N$ | $g_{11}^*$ | 0.6 |
| $G_{12}$ | 0 | $C^{-2}m^4N$ | $g_{12}^*$ | 0 |
| $G_{44}$ | $4.734 \times 10^{-10}$ | $C^{-2}m^4N$ | $g_{44}^*$ | 0.3 |
| $G_{44}'$ | $4.734 \times 10^{-10}$ | $C^{-2}m^4N$ | $g_{44}'^*$ | 0.3 |
| $C_{11}$ | $3.02 \times 10^{11}$ | $N \cdot m^{-2}$ | $C_{11}^*$ | 1929 |
| $C_{12}$ | $1.62 \times 10^{11}$ | $N \cdot m^{-2}$ | $C_{12}^*$ | 1035 |
| $C_{44}$ | $6.8 \times 10^{10}$ | $N \cdot m^{-2}$ | $C_{44}^*$ | 434 |
| $Q_{11}$ | 0.032 | $C^{-2}m^4$ | $q_{11}^*$ | 0.013 |
| $Q_{12}$ | -0.016 | $C^{-2}m^4$ | $q_{12}^*$ | -0.006 |
| $Q_{44}$ | 0.06 | $C^{-2}m^4$ | $q_{44}^*$ | 0.048 |
| $\varepsilon_r$ | 70 | - | - | - |

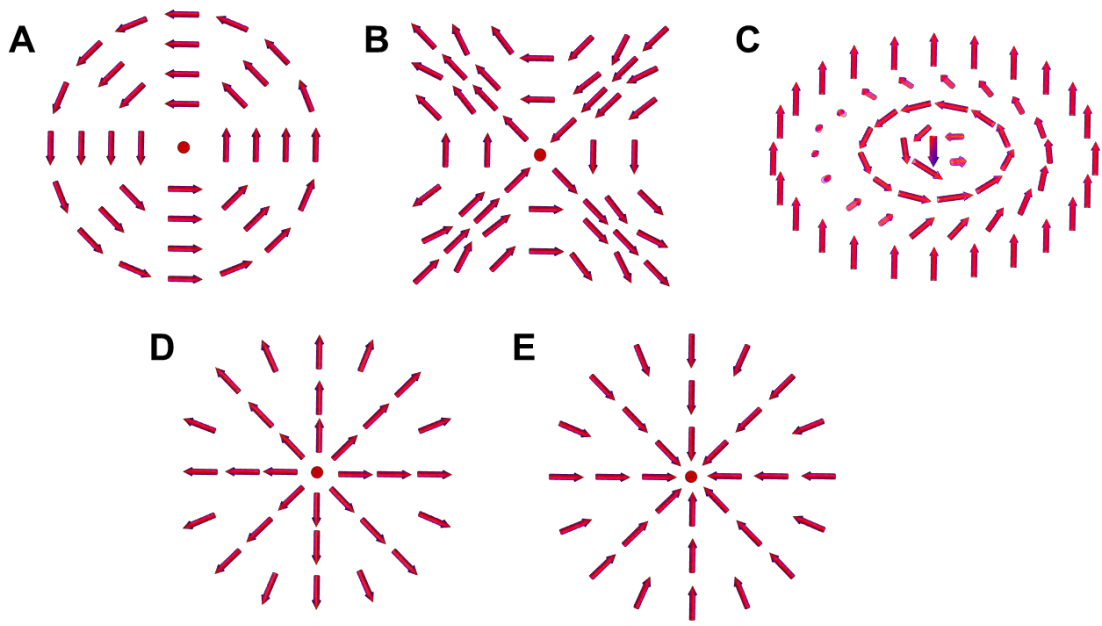

**fig. S1. Schematic diagrams for different 1D topological defects.** (**A**) Flux closure vortex domain. (**B**) Anti-vortex domain. (**C**) Skyrmion domain. (**D**) Divergent center domain. (**E**) Convergent center domain.

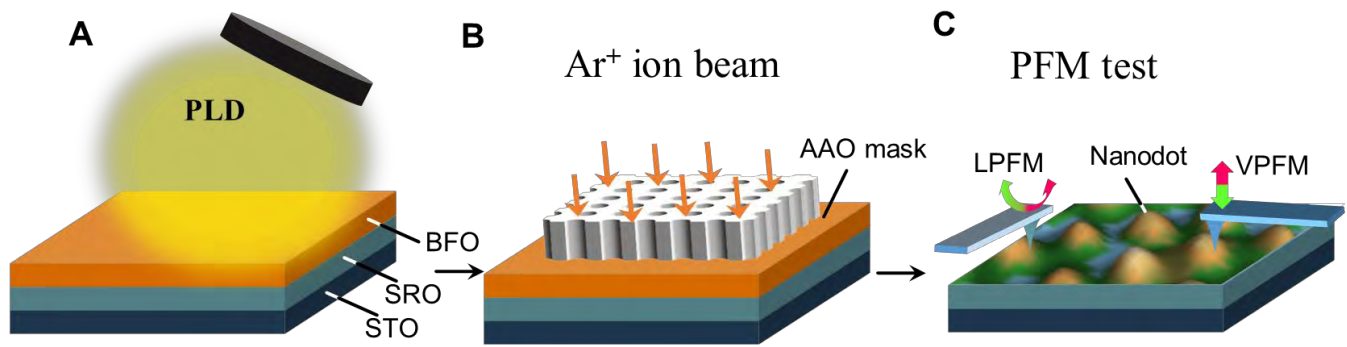

**fig. S2. Schematic procedures for the fabrication and PFM characterization of BFO nanodots array sample.** (**A**) Pulsed laser deposition of the epitaxial BFO/SRO heterostructured film on STO substrate. (**B**) Ar$^+$ Ion beam etching of the BFO film with a sacrificed AAO template to form the nanodots. (**C**) Vertical and lateral PFM characterizations.

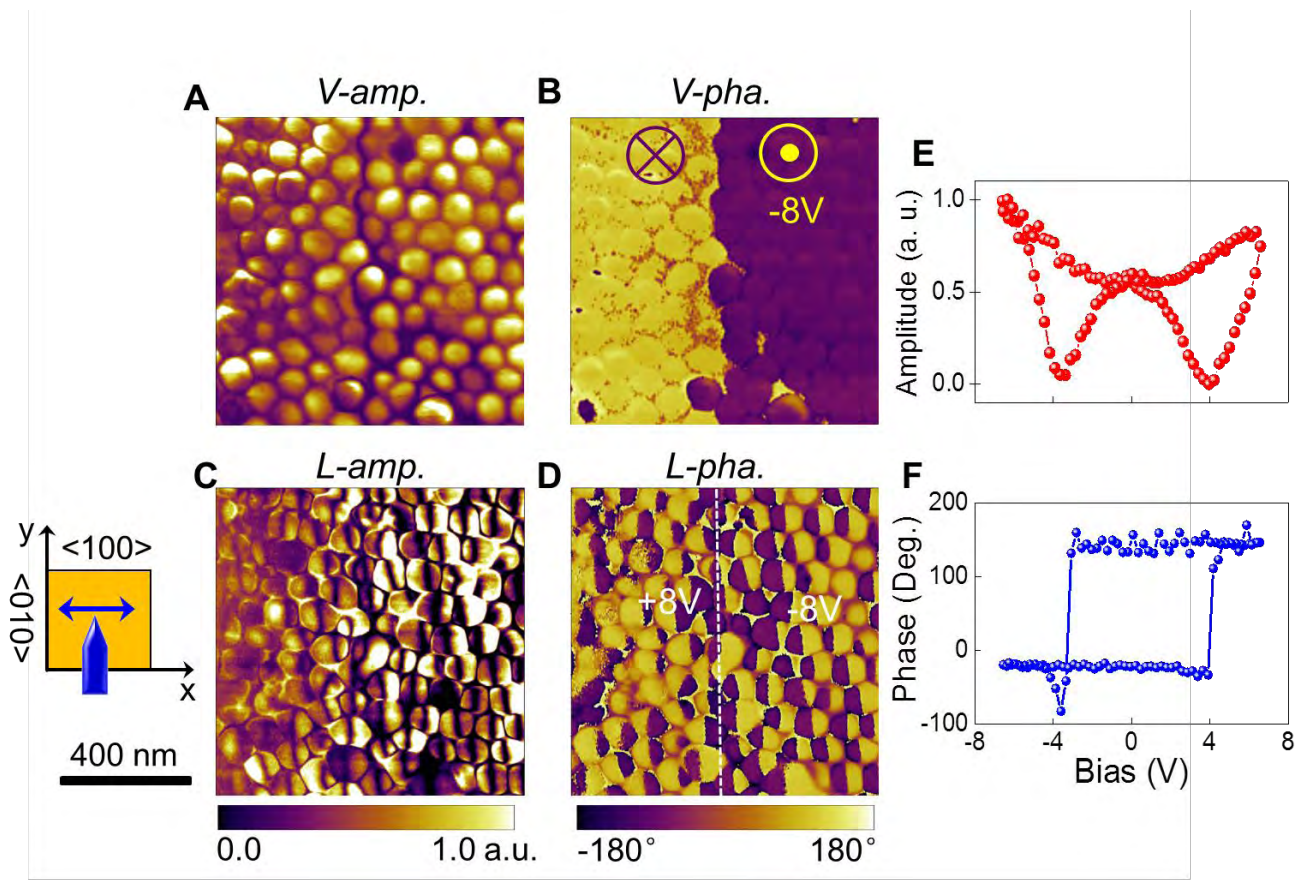

**fig. S3. PFM images for the nanodot array after poling by scanning bias voltages of ±8 V.** (**A** to **D**) The vertical PFM amplitude (**A**) and phase (**B**) images, as well their corresponding lateral PFM amplitude (**C**) and phase (**D**) images for the nandot array. (**E** and **F**) The piezoresponse butterfly amplitude (**E**) and square phase (**F**) loops, respectively, for a selected nanodot.

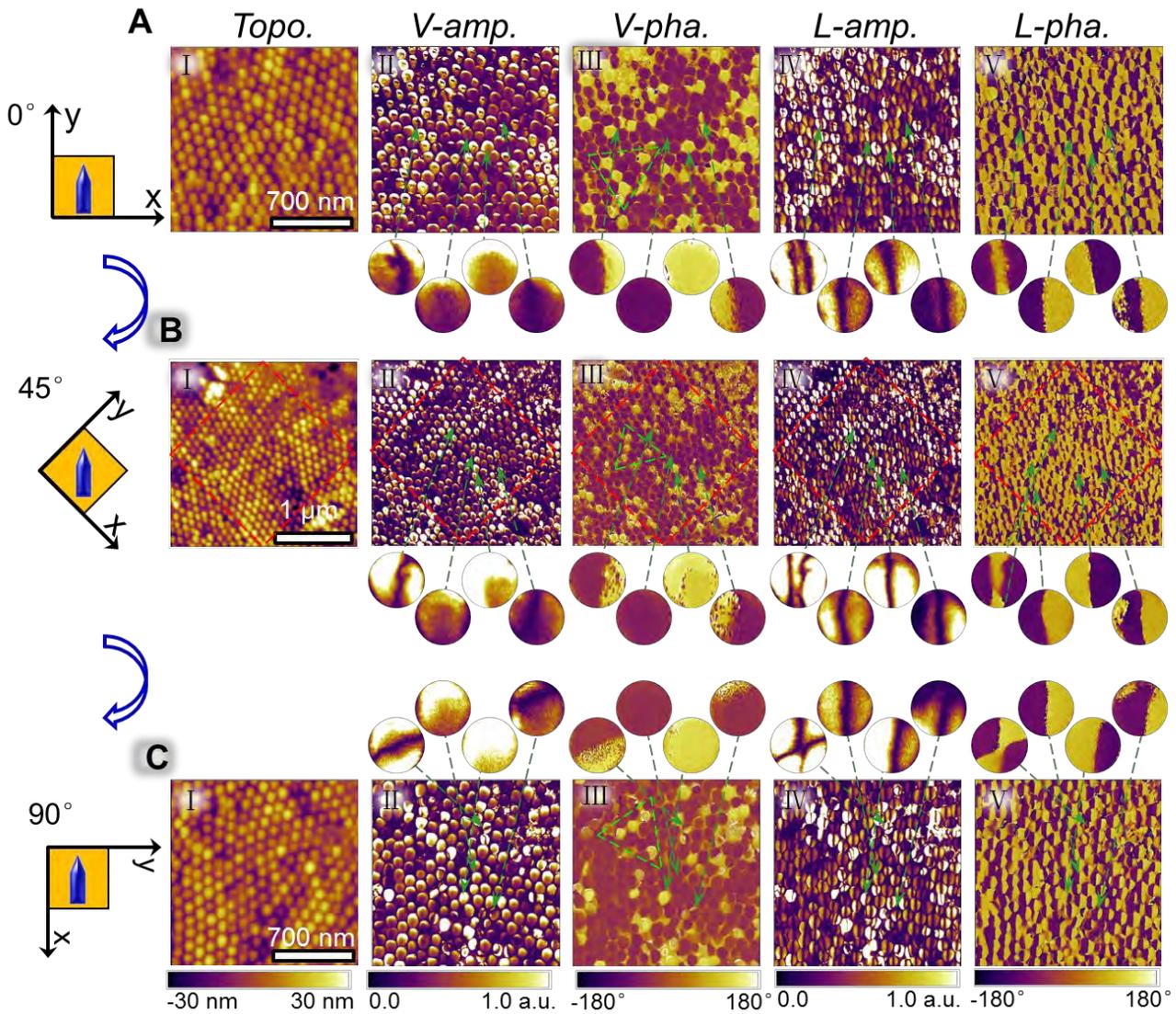

**fig. S4. 3D PFM images for a nanodot array**. (**A** to **C**) The topographic, PFM vertical amplitude and phase, and lateral amplitude and phase images, for three different sample rotation angles: 0º (**A**), 45º (**B**), and 90º (**C**). The inset zoom-in single dots PFM images in the gap between **A**, **B** and **B**, **C** illustrate the different aspects of PFM contrast for some frequently observed domain structures in the nandots. The square region surround by red doted lines in b is correspondent to the same region as those of the whole image in a and b. The green triangles in *V-pha* images in a-c indicate the same region after different in-plane rotations.

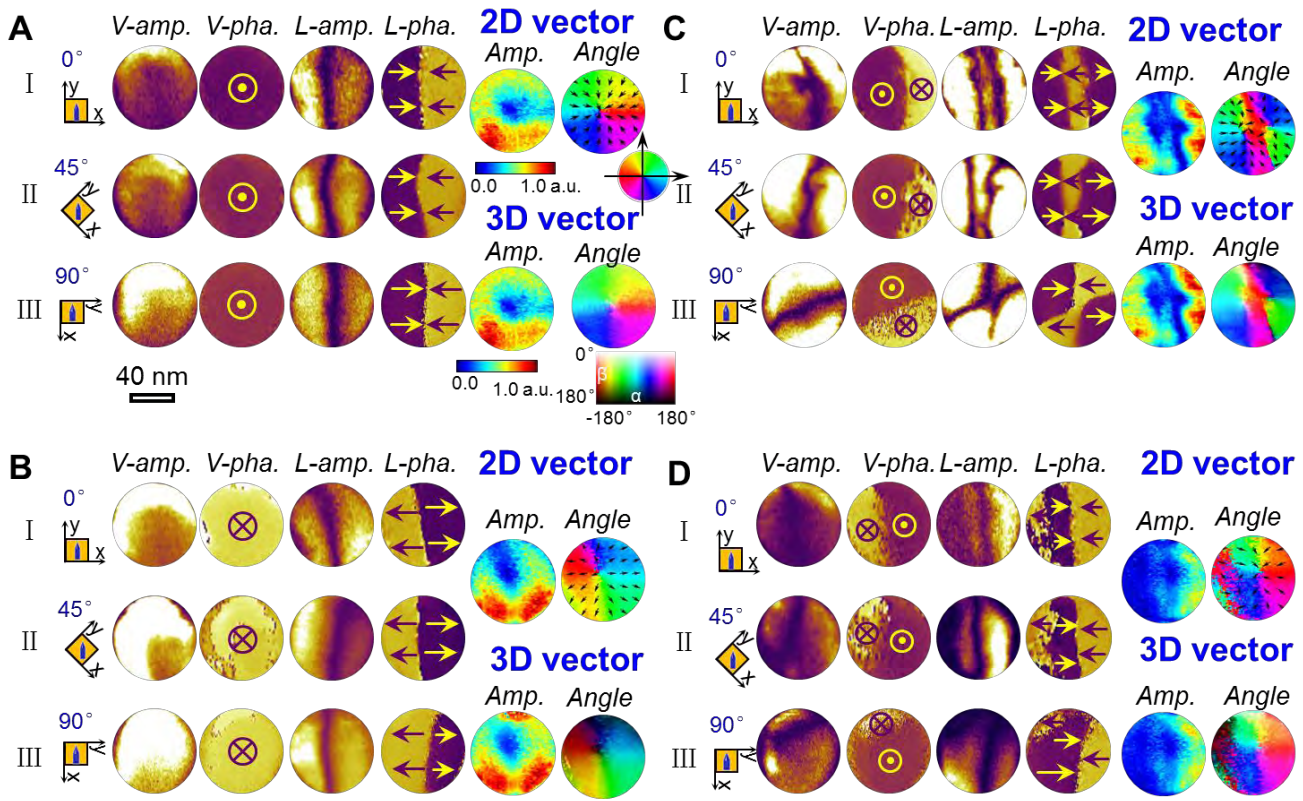

**fig. S5. Single dot PFM images for four typical topologic domains.** (**A** to **D**) The detailed PFM images micrographs, along with vector contour maps and schematic diagrams for the different types of topologic domain structures, including center-convergent domain (**A**), center- divergent domain (**B**), double-center domain (**C**), and reverse-double center domains (**D**) structures. In each domain states, PFM images were taken with sample clockwise ration for 0º (**I**), 45º (**II**), and 90º (**III**), and both vertical and lateral PFM (amplitude and phase) images are shown in every individual rotation angle. The color contour maps present the 2D lateral vector and 3D vector maps derived from the analysis of the corresponding raw PFM data, for the four different domain structures.

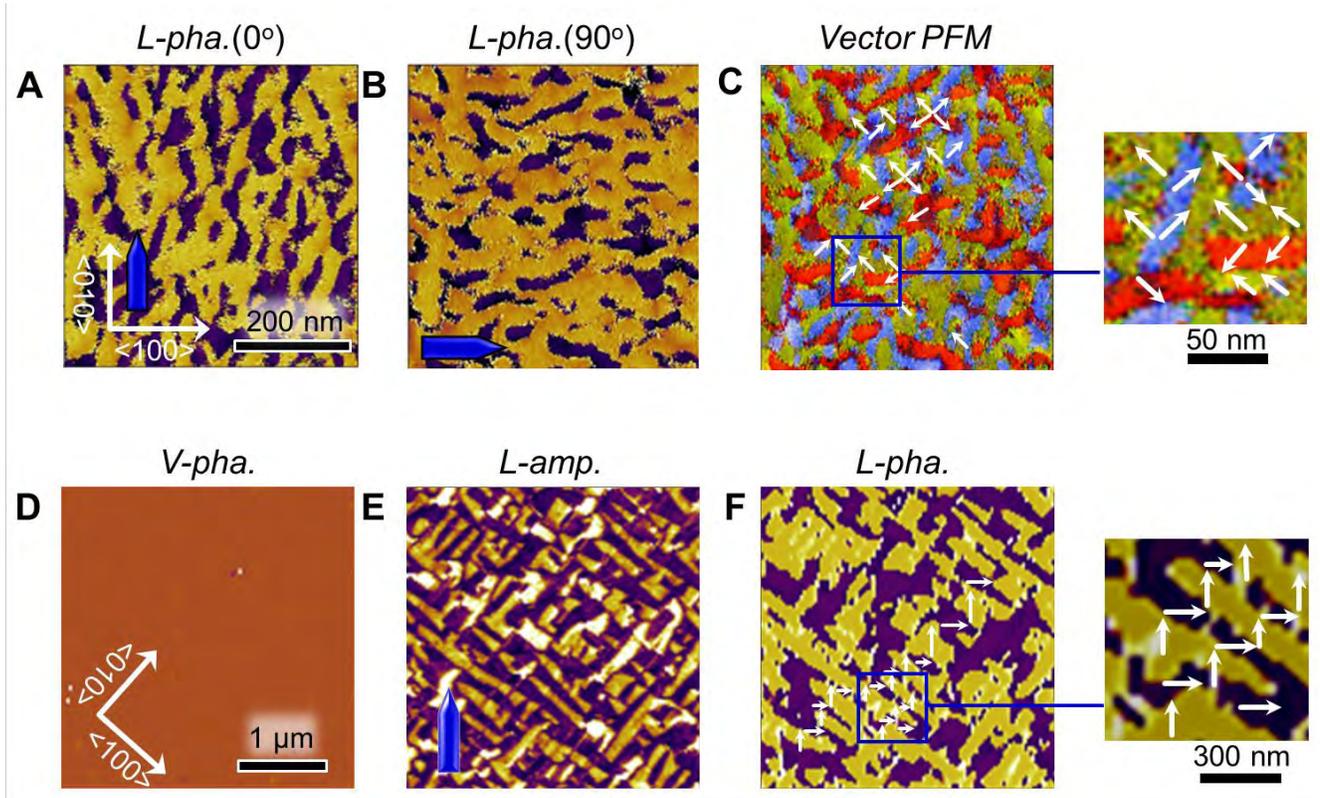

**fig. S6. A comparison of domain structures between an oxygen deficient BFO film and a less oxygen deficient film.** (**A** to **C**) Lateral PFM phase images and domain structures for the oxygen deficient BFO film deposited at a low oxygen pressure of 3 Pa, including lateral phase image for the same area scanned with cantilever along <010> direction (**A**), along <100> direction (**B**), and vector PFM domain images (**C**) derived from the combination of a and b. The different color contrasts in **C** present the four different in-plane polarization orientations, indicating head-to-head or tail-to-tail charged domain walls are predominant in the oxygen deficient film. (**D** to **F**) the PFM vertical phase (**D**), lateral amplitude (**E**), and lateral phase (**F**) images for a less oxygen deficient BFO film, which was deposited at a relatively high oxygen pressure of 13 Pa and subsequently cooled down to room temperature at 5000 Pa. The cantilever is along the <110> direction. The typical stripe domains in F indicate the film is dominant 71° domain walls, free of head-to-head or tail-to-tail charged domain walls. The inset magnified vector PFM image in **C** clearly indicates the dominance of charged domain walls in the oxygen deficient film, and the inset magnified PFM phase image in e indicates the predominance of 71° tail-to-head charge free domain walls in the less oxygen deficient film.

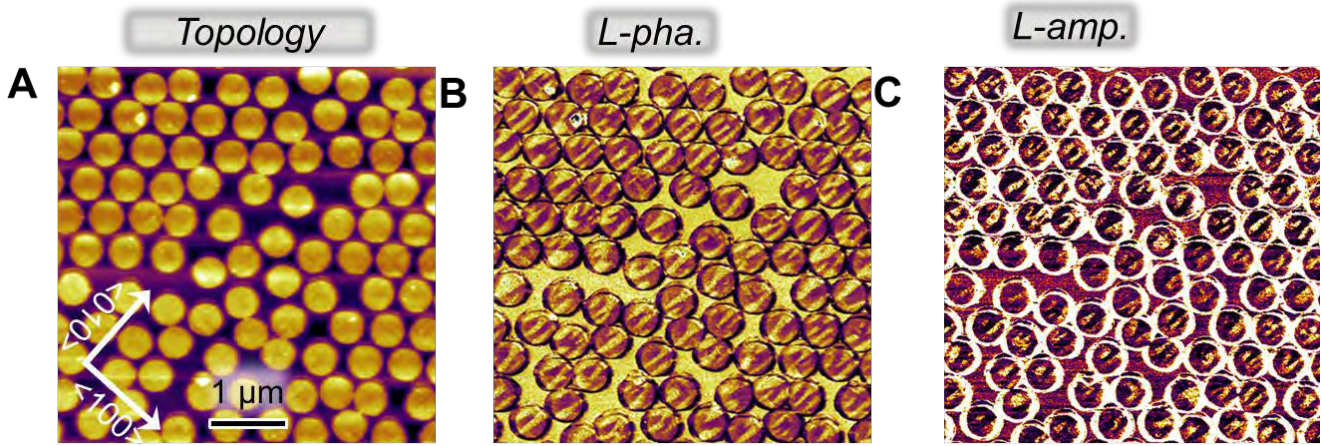

**fig. S7. Stripe domain structures in nanodots derived from a less oxygen deficient BFO film.** (**A** to **C**) The topology (**A**), phase (**B**), and amplitude (**C**) micrographs for the nandots array derived from a less oxygen deficient film by mask assist ion beam etching. Most of the nanodots exhibit well defined 71° stripe pattern domain, which is rather similar to that of their parent BFO film prior to ion beam etching.

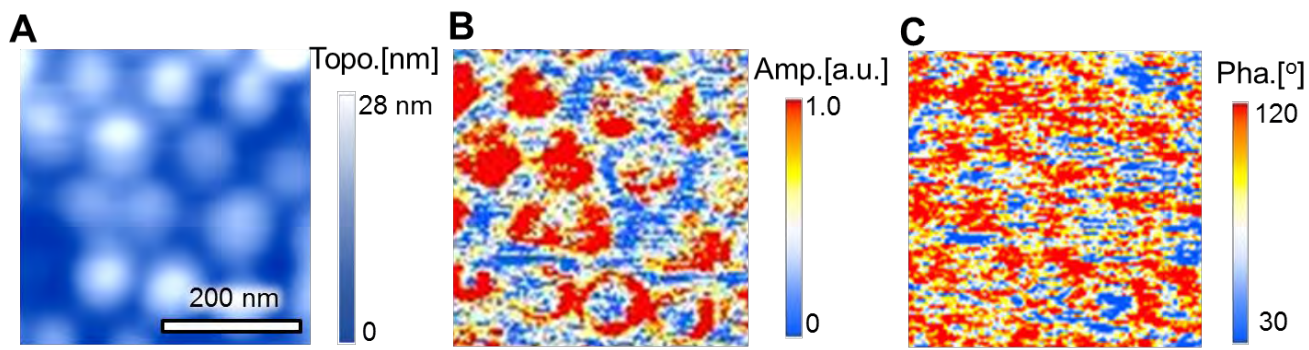

**fig. S8. Scanning thermal-ionic microscopy (STIM) images of a nanodot array in this work.** (**A** to **C**), The topology (**A**), amplitude (**B**), and phase (**C**) micrographs for the nandot array. The bright contrast in amplitude map reflects the accumulation of ionic charge carriers (oxygen vacancies) carrying positive charge locating in the centers or outer edges (or bottom) of the dots.